# Low-dispersion low-loss dielectric gratings for efficient ultrafast laser pulse compression at high average powers

David A. Alessi,* Hoang T. Nguyen, Jerald A. Britten, Paul A. Rosso, and Constantin Haefner

*Lawrence Livermore National Laboratory, Livermore, California, 94550, USA*

*Corresponding author: alessi2@llnl.gov

## ABSTRACT

We have developed low-dispersion (1480 l/mm), resonance-free, diffraction gratings made of dielectric materials resistant to femtosecond laser damage ($SiO_2/HfO_2$). A 14 cm diameter sample was fabricated resulting in a mean diffraction efficiency of 99.1% at $\lambda$ = 810 nm with 0.4% uniformity using equipment which can fabricate gratings up to 1 m diagonal. The implementation of these gratings in the compression of 30 fs pulses in an out-of-plane geometry can result in compressor efficiencies of ~95%. The measured laser absorption is 500x lower than current ultrafast petawatt-class compressor gratings which will enable a substantial increase in average power handling capabilities of these laser systems.

# 1. INTRODUCTION

There is growing interest in the development of intense lasers which operate at high repetition rates for scientific and industrial applications [1-5]. Many of these applications require petawatt-class peak powers having kilowatt to megawatt average powers. Compressors for these high peak power laser systems with sub-150fs pulse durations utilize gold gratings because of their broad spectral bandwidth [6] but are typically limited to average powers ($P_{avg}$) of several watts due to thermally induced deformation from absorption [7,8] and are fundamentally limited to efficiencies of ~80%. The repetition rates of petawatt-class ultrafast systems has been steadily increasing with the BELLA laser [9] operating at 1 Hz ($P_{avg}$ = 40 W), and a 3.3 Hz Ti:sapphire system recently commissioned at CSU ($P_{avg}$ = 85 W) [10]. LLNL's diode-pumped High-repetition-rate Advanced Petawatt Laser System (HAPLS) [11,12] installed in ELI Beamlines [5] is designed to produce petawatt pulses at 10 Hz ($P_{avg}$ = 300 W). To address the increase in average power requirements, LLNL has recently developed Gold Coated Dielectric Ridge (GCDR) gratings [13] which have improved efficiency and uniformity as compared to previous high average power class gratings consisting of gold coated etched substrates. Benchmarked thermo-mechanical modeling predicts that photoresist-free gold gratings with low thermal-expansion substrates combined with active edge cooling can enable compression of petawatt pulses with average powers above 500 W [14].

Multi-Layer Dielectric (MLD) gratings were developed for their high efficiency and high laser damage threshold. The first meter-class MLD gratings were produced at LLNL [15] and implemented on high-energy short-pulse petawatt-class systems operating at single-shot repetition rates. In addition, MLD gratings have 2-3 orders of magnitude less absorption than gold and are ideal for addressing the wide gap between gold grating-based compressor average power limits and the need for applications. Reflection-mode MLD gratings are commercially available at ~1 µm wavelengths for 30-40 nm bandwidths [16, 17], and designs for or even broader bandwidths have been demonstrated [18, 19]. However, producing MLD gratings suitable for high peak power Ti:Sapphire laser systems has been a significant challenge. Guided Mode Resonances (GMRs) [20] can easily occur within broad-bandwidth MLD gratings resulting in absorption, laser damage, and degraded pre-pulse contrast. These can be eliminated by reducing the grating period [21] and has resulted in several MLD gratings for λ = 800nm with high line densities (1700-1900 l/mm) [21-25]. However, to our knowledge, no MLD gratings have been utilized in the pulse compression of broad bandwidth Ti:Sapphire laser systems producing peak powers above 0.1 PW. This is due to the fact that low line densities are required for both dispersion management [26] and system architecture design of high intensity laser systems. Recently a broad-bandwidth grating design with 1580 l/mm was proposed using Gires-Tournois resembling cavity layers [27]. However, this design is not optimal for high energy pulse compression due to the fact that $TiO_2$, a material with low femtosecond Laser Induce Damage Threshold (LIDT) [28], is used in both the grating pillar and the multilayer stack. The utilization of high LIDT materials is required to maximize energy density and achieve maximum intensities on target with limited grating size. In this work we have designed and fabricated low dispersion (1480 l/mm) gratings which are scalable to meter sizes to enable high efficiency petawatt-class pulse compression of λ = 800nm pules to sub-20 fs durations at high average powers.

## 2. GRATING DESIGN

Grating performance modeling was carried out using both the Gsolver [29], and LightTrans VirtualLab with grating toolbox [30] commercial software packages. Gsolver utilizes the full vector treatment of the rigorous coupled wave analysis while the VirtualLab grating toolbox can also use the numerical fourier modal method. The grating line density of 1480 l/mm was selected based on the system design of a high energy Ti:sapphire laser system and also corresponds to the value for operational mult-petawatt laser systems [31, 32]. It is well known that femtosecond LIDT in MLD gratings is strongly dependent on the electric-field in the grating ridges [6], and as a result, the grating ridge material was selected to be $SiO_2$ due to its high femtosecond LIDT compared to other coating materials [28]. The coating stack materials ($SiO_2$/$HfO_2$) were also chosen because of their high LIDT at femtosecond pulse durations. With the line density and materials fixed, the grating design was modified to produce the broadest bandwidth with no GMRs. The final grating design is shown in Fig 1a and consists of a 42 layer modified quarter wave stack with a total thickness (D) of 5.83 µm with rectangular pillars etched into most of the top layer. The grating pillar structure has a 22% duty cycle (f) and height (H) of 650 nm. The grating diffraction efficiency as a function of wavelength is shown as the solid blue line in Fig. 1b. The laser polarization is transverse electric (s-polarized) and the incidence angle is the Littrow angle for λ = 810 nm (36.8° with respect to the normal). This is compared to the measured efficiency (solid orange line) of recently developed high average power compatible GCDR gratings [13] under common use conditions (56° incidence angle).

We calculated the MLD grating diffraction efficiency as a function of wavelength for a wide range of input angles. These angles (θ,ϕ) are defined in Fig. 2a where θ is the traditionally defined angle of incidence in the plane normal to the grooves, and ϕ is the angle from the input vector to this plane (out of plane angle). The efficiency as a function of wavelength for different input angles is shown in Fig. 2b. The best performance occurs at θ = 36.8°, ϕ = 0° (blue line) and degrades significantly while the input angle (θ) is changed only 4° from Littrow angle (red line). A four-grating compressor operating at this input angle to separate the input and output beams would have an overall efficiency of ~50%. However, a 4° change in ϕ only results in a slight reduction of the efficiency (green line). Fig. 2c shows the diffraction efficiency as a function of angle (θ,ϕ) for λ = 810 nm with s-polarization. These simulations show that the efficiency decreases ~4x faster with increasing θ than with increasing the out of plane angle ϕ. In order to separate input and output beams in a compressor, these gratings should operate with θ = 36.8° and an out-of-plane angle. This type of out-of-plane compressor has been demonstrated using gold gratings for broad-bandwidth λ = 800 nm lasers [33].

## 3. MEASURED GRATING PERFORMANCE

To reduce this design to practice, grating samples were fabricated up to 14 cm in diameter. The grating ridges were patterned and etched into the top $SiO_2$ layer in-house by holographic lithography followed by reactive ion-beam etching [34] using equipment which produces MLD gratings with sizes up to 1 m diagonal. Fig. 3a shows a map of the spatial variation of the diffraction efficiency at λ = 810 nm with θ = 36.8°, ϕ = 0 incidence angle and s-polarization measured with 1.5 mm resolution. A histogram of the nominal 100 mm x 100 mm region of interest is shown in Fig 3b. The diffraction efficiency has a mean value of 99.1% at this wavelength, and the standard deviation is 0.4% of the mean. The measured diffraction efficiency as a function of wavelength for θ = 36.8°, ϕ = 0° incidence angle is shown as the red circles in Fig. 1b with an uncertainty of < 0.1%. It is reasonable to conclude that these gratings are GMR free due to the excellent agreement between the measured and modeled diffraction efficiency as a function of wavelength. Optical absorption measurements were performed at λ = 830 nm using photo-thermal common path interferometry at Stanford Photo Thermal Solutions where the average surface absorption of 3 samples varied from 62-108 ppm. These samples were coated using electron beam deposition and there is potential to reduce the grating absorption further by using ion beam sputtering, a coating process which is known to produce low-absorption coatings [35].

Preliminary laser damage testing was performed in-house using a 10 Hz broad-bandwidth Ti:sapphire laser. Laser pulses with ~40 nm bandwidth were compressed to 80 fs and directed into a laser damage test station (in air). The damage test station utilized is based on that described by Negres [36] in which the energy and reference focal spot of each laser pulse delivered to the target is captured. A 10-site r-on-1 measurement of an MLD grating using an incidence angle of 37° and s-polarization resulted in a 50% damage probability fluence of 0.29 J/cm$^2$ projected in the plane normal to the beam. For comparison, the same measurement conditions were performed using a gold grating which resulted in a 50% damage probability fluence of 0.34 J/cm$^2$. The 50% damage probability fluence is intended to compare performance between samples, not to provide an operational limit. Damage testing was also performed on the same gratings using the stretched pulse (400 ps). At this pulse duration, damage measurements were obtained on the gold grating but not the MLD grating due to limitations in the laser fluence. From this, we can only conclude that the laser damage resistance of the MLD grating is at least four times higher than the gold grating.

## 4. ESTIMATED COMPRESSOR PERFORMANCE

We estimate the performance of a pulse compressor utilizing these gratings using the measured wavelength-dependent diffraction efficiency of a single grating. First, the grating efficiency at each wavelength is taken to the 4th power to estimate the overall compressor efficiency. Second, an input laser pulse spectrum is generated using both gaussian and 20$^{th}$ order supergaussian shapes. Third, the compressor output spectral amplitude is determined using the estimated compressor spectral transmission function. Fourth, the overall compressor efficiency is determined by taking the ratio of the integrated output spectrum divided by the input spectrum. Fifth, the transform limit pulse duration of the compressor output is calculated from the inverse fast fourier transform of the output spectrum. Fig. 4 shows the compressor efficiency and output pulse duration as a function of input laser FWHM bandwidth. For a pulse with a 20$^{th}$ order supergaussian shape having a FWHM bandwidth of 60 nm, a compressor with these MLD gratings could support an output pulse duration of 30 fs with an overall compressor efficiency of 93%. This is ~2.5x less loss than the current state-of-the-art compressors for this pulse duration. The compressor spectral transmission can support pulse durations as short as 15 fs. At 18 fs, the compressor efficiency is similar to a GCDR grating based compressor but withstands two orders of magnitude higher average powers due to the reduced absorption.

We have estimated the performance of these gratings when used for the compression of high average power ultrafast lasers using measured absorption data and thermo-mechanical modeling [14]. Fig. 5 shows the steady state surface deflection and temperature from finite element modeling calculations of the input compressor grating of a petawatt-class laser with 40 J of energy per pulse as a function of laser average power. These simulations are for a single 0.27 m tall by 0.5 m wide grating with a beam footprint of 0.39 m wide x 0.21 m tall where the absorption is 62 ppm. The red lines are the grating peak temperature and the blue lines are the grating surface peak-to-valley deflection. The substrate used is an ultra-low thermal expansion glass substrate which is either un-cooled (solid lines) or actively cooled from the top and bottom surfaces (dashed lines). A yellow line is drawn at 80 nm corresponding to a reflected wave front distortion of 0.2λ, a value chosen to compare relative performance of different grating technologies. These estimates predict that this particular deformation limit is reached at $P_{av}$ = 76 kW (1.9 kHz at 40 J/pulse) for passive cooling, which is more than 500x higher than current broad-bandwidth petawatt-class compressor technologies. Active cooling would result in an increase to Pav = 320 kW (8 kHz at 40 J/pulse). Additionally, the thermal performance could improve by cooling from the back surface or developing coatings with lower absorption.

## 5. CONCLUSION

In conclusion, we have designed and fabricated the first MLD gratings suitable for ultrafast laser systems producing petawatt-class peak powers. This was accomplished by fabrication of gratings with multiple specifications simultaneously: low dispersion, scalable to meter sizes, GMR-free, broad-bandwidth, and consisting of materials resistant to laser damage ($SiO_2/HfO_2$). The grating line density of 1480 l/mm was chosen from the laser system design and corresponds to the value used in multi-petawatt facilities worldwide. A 14 cm diameter grating was produced using equipment that fabricates MLD gratings up to 1 m diagonal. This sample produced high average efficiency (99.1%) and high spatial uniformity (0.4%) at the center wavelength. Analysis of the measured diffraction efficiency shows that these gratings support high energy compression to pulse durations below 20 fs and these mid-scale gratings could produce 30 fs pulses with a compressor efficiency of 93%, having 2.5x less loss than existing state-of-the art. Obtaining this performance relies on operating the gratings with a near-littrow out-of-plane architecture. The optical absorption of sample gratings was measured to be as low as 62 ppm, which is 500x lower than current state-of-the art petawatt-class ultrafast compressor gratings. We predict that these gratings can enable pulse compressors for petawatt-class systems operating at repetition rates as high as 8kHz when combined with active cooling.

**Funding.**

This work was performed under the auspices of the U.S. Department of Energy by Lawrence Livermore National Laboratory under Contract DE-AC52-07NA27344; Laboratory Directed Research and Development (LDRD) Program under tracking code 14-ERD-084

**Acknowledgment.**

We thank E. Sistrunk and T. Spinka for useful discussions.

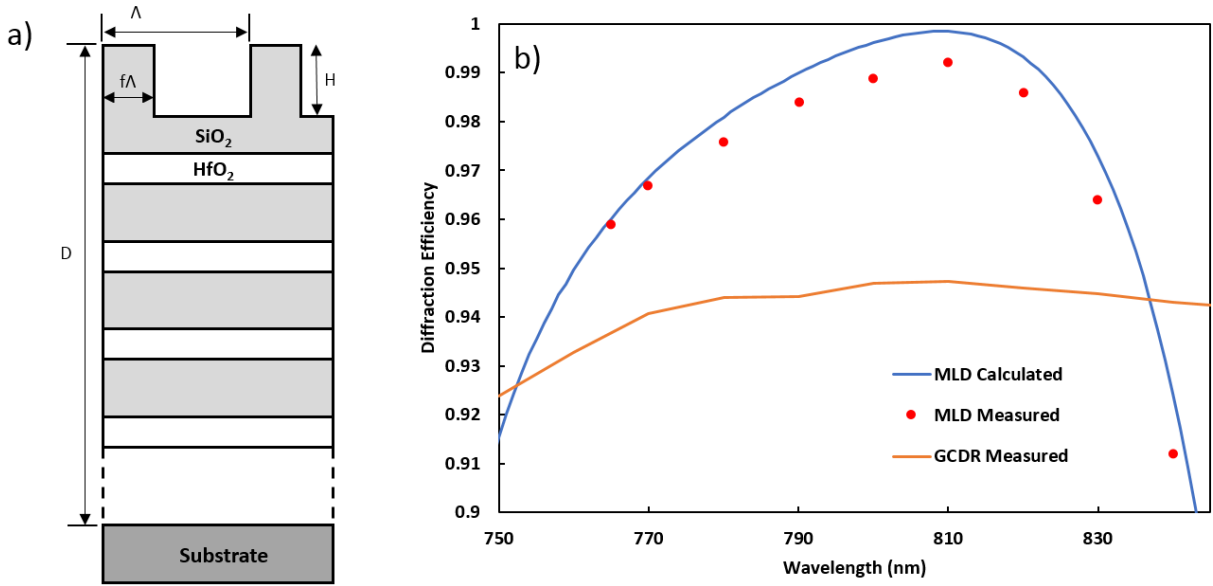

Fig. 1. (a) Schematic diagram of the final MLD grating design with grating period (Λ), duty cycle (f), pillar height (H), and a 42 layer multilayer stack with total thickness (D). (b) Diffraction efficiency as a function of wavelength for 1480 l/mm gratings. Measured (red dots) and calculated (blue line) efficiencies for MLD gratings are shown for 36.8° incidence angle and s-polarization. Measured values for gold coated dielectric ridge (GCDR) gratings are shown for a 56° incidence angle and p-polarization (orange line).

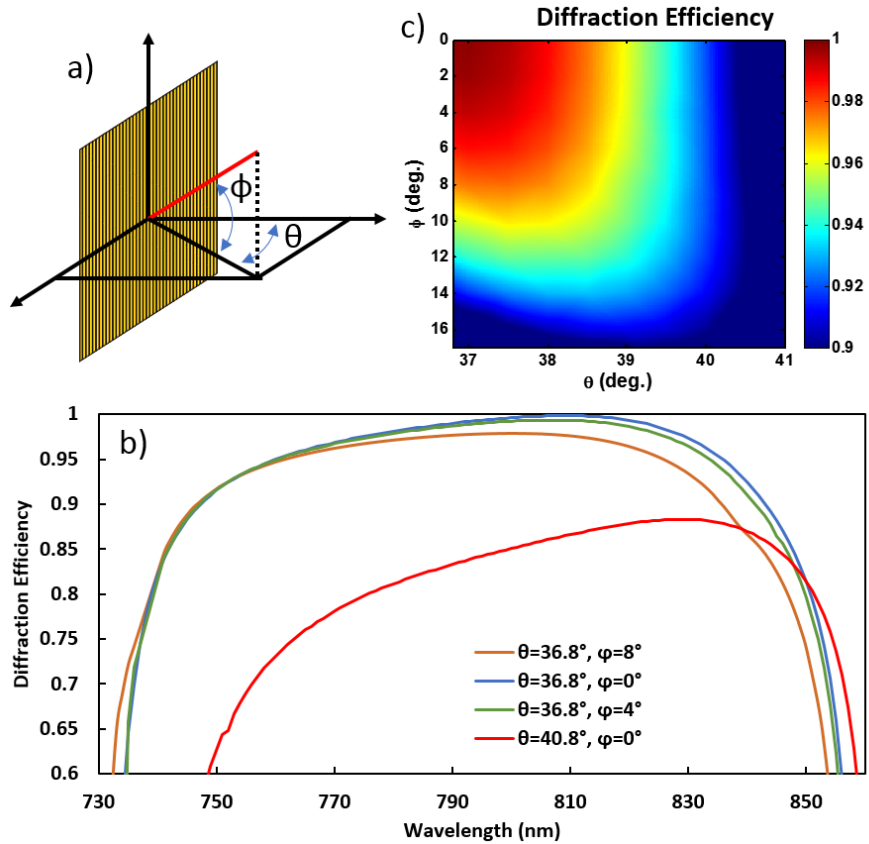

Fig. 2. (a) Input angles θ and ɸ defined with respect to grating surface showing direction of grooves. (b) Simulated diffraction efficiency as a function of wavelength for θ = 36.8°, ɸ = 0° (blue); θ = 40.8°, ɸ = 0° (red); θ = 36.8°, ɸ = 4° (green); θ = 36.8°, ɸ = 8° (orange). (c) Simulated diffraction efficiency at λ = 810 nm as a function of both θ and ɸ.

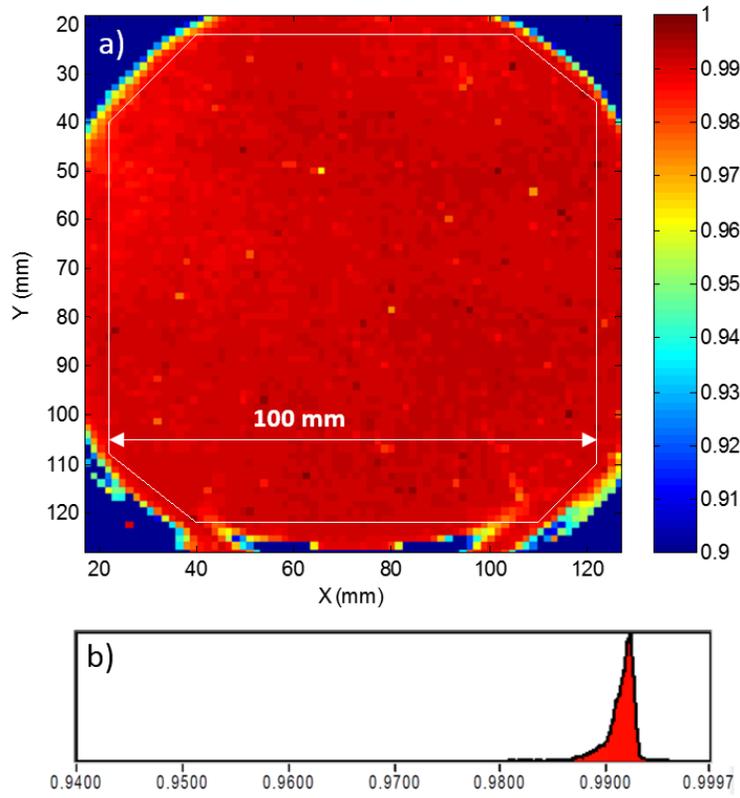

Fig. 3. (a) Diffraction efficiency of a 14 cm diameter 1480 l/mm MLD grating measured as a function of position for λ = 810nm incident at 36.8° with s-polarization. (b) Histogram of efficiency within the region of interest (white boundary) for which the mean is 99.1% and standard deviation is 0.4% of the mean.

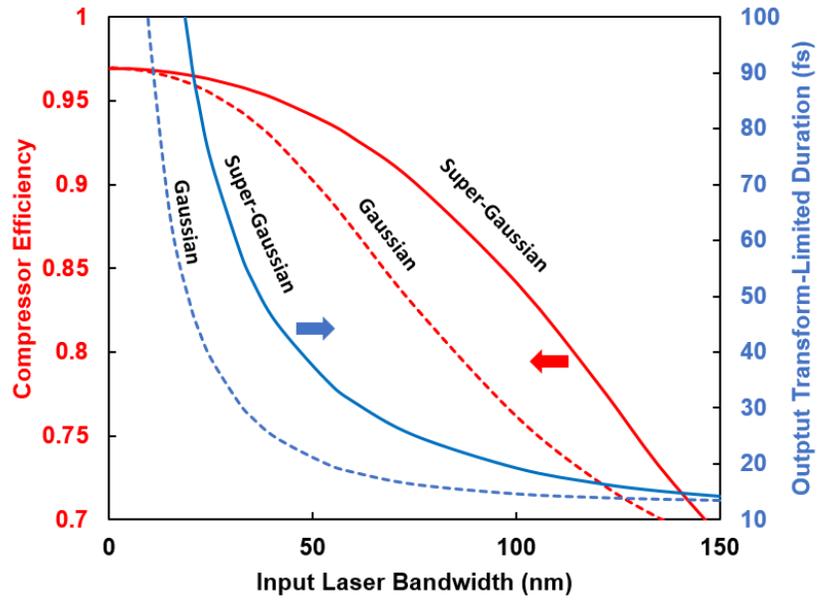

Fig. 4. Simulated compressor efficiency (red lines) and laser output transform limited pulse duration (blue lines) as a function of input laser FWHM bandwidth based on measured grating efficiency data. Calculations are performed for a gaussian (dotted lines) and a 20$^{th}$ order supergaussian (solid lines) shaped spectrum.

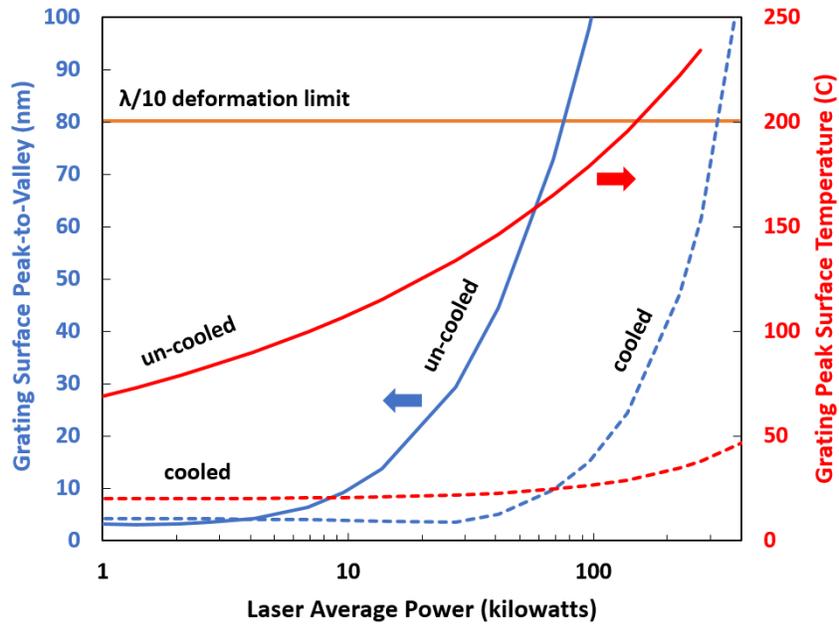

Fig. 5. Simulated steady-state grating surface peak-to-valley (blue lines) and peak temperature (red lines) as a function of laser average power for a petawatt-scale compressor grating in vacuum using the measured grating absorption of 62 ppm. Solid lines are for the un-cooled case and dashed lines are results for active substrate cooling.